\documentclass[conference]{IEEEtran}
\usepackage{pgfplots} 
\usepackage{tikz}
\usetikzlibrary{positioning, shapes, shadows, arrows}
\usepackage{graphicx,cite,multirow,rotating}
\usepackage{amssymb}
\usepackage{amsthm}
\usepackage{amsfonts}
\usepackage{yfonts}
\usepackage{psfrag}
\usepackage{amsfonts}
\usepackage[utf8]{inputenc}
\usepackage[linesnumbered,ruled]{algorithm2e}
\usepackage{bbm}
\usepackage{dsfont}
\usepackage{subfigure}

\def\vec{{\rm vec}}
\def\rE{{\rm E}}

\def\arg{{\rm arg}}

\def\d={{\stackrel {\rm def}{=}}}

\newcommand{\beq}{\begin{equation}}

\newcommand{\eeq}{\end{equation}}
\newcommand{\bqy}{\begin{eqnarray}}
\newcommand{\eqy}{\end{eqnarray}}

\newcommand{\xx}[1]{{\mathbf x}}
\newcommand{\yy}[1]{{\mathbf{y}}}
\newcommand{\XX}[1]{{\mathbf{X}}}
\newcommand{\YY}[1]{{\mathbf{Y}}}
\newcommand{\RR}[1]{{\mathbf{R}}}
\newcommand{\HH}[1]{{\mathcal{H}}}

\def\BD{\begin{displaymath}}
\def\BEA{\begin{eqnarray}}
\def\BEAs{\begin{eqnarray*}}
\def\ED{\end{displaymath}}
\def\EE{\end{equation}}
\def\EEA{\end{eqnarray}}
\def\EEAs{\end{eqnarray*}}

\def\bA{\textbf{A}}

\def\bA{{\bf A}}

\def\bB{{\bf B}}

\def\bC{{\bf C}}

\def\be{{\bf e}}

\def\bH{{\bf H}}

\def\bI{{\bf I}}

\def\bl{{\bf l}}

\def\bQ{{\bf Q}}

\def\bR{{\bf R}}

\def\bs{{\bf s}}

\def\bW{{\bf W}}
\def\bw{{\bf w}}
\def\bX{{\bf X}}
\def\bx{{\bf x}}
\def\bY{{\bf Y}}
\def\by{{\bf y}}

\def\rE{{\rm E}}
\def\re{{\rm e}}

\def\rP{{\rm P}}

\def\rT{{\rm T}}

\def\b_eta{\mbox{\boldmath $\eta$}}

\def\bnu{\mbox{\boldmat $\nu$}}

\def\diag{{\rm diag}}
\def\vec{{\rm vec}}

\def\L2{L^{2}[-{\pi \over 2} , {\pi \over 2}]}

\def\bnu{\bnu}

\usepackage{setspace}
\usepackage{color}
\usepackage{amsmath,stackrel,dsfont}
\usepackage{amssymb,hyperref,stmaryrd}

{}

\def\b1{{\mathbf{1}}}
\def\rE{{\rm E}}

\title{  { A Complexity Efficient DMT-Optimal Tree Pruning Based Sphere Decoding}}

\author{Mohammad~Neinavaie,~Mostafa~Derakhtian, Negar~Daryanavardan and Sergiy~A.~Vorobyov,~{\it Fellow},~IEEE
}
\begin{document}

\maketitle

\begin{abstract}
	We present a diversity multiplexing tradeoff (DMT) optimal tree pruning sphere decoding algorithm which visits merely a single branch of the search tree of the sphere decoding (SD) algorithm, while maintaining the DMT optimality at high signal to noise ratio (SNR) regime. The search tree of the sphere decoding algorithm is pruned via intersecting one dimensional spheres with the hypersphere of the SD algorithm, and the radii are chosen to guarantee the DMT optimality. In contrast to the conventional DMT optimal SD algorithm, which is known to have a polynomial complexity at high SNR regime, we show that the proposed method achieves the DMT optimality by solely visiting a single branch of the search tree at high SNR regime. The simulation results are corroborated with the claimed characteristics of the algorithm in two different scenarios.
\end{abstract}
\section{Introduction} \label{radarapp}
The problem of {d}ecoding  information symbols that are drawn from a finite set leads to the integer least square (LS) problem. The integer LS problem arises in many applications, e.g. the decoding of high rate space-time codes \cite{new1}. The optimum solution for the integer LS problem can be obtained by the maximum likelihood (ML) detector. A brute-force search is a computationally inefficient way of solving the ML detection problem specially for a large number of lattice points. Thus, in the past decade, the sphere decoding (SD) algorithm has been frequently employed and studied in order to reduce the computational complexity of exact or near ML decoding in the integer LS problem \cite{new2}.

Analysis of the computational complexity of the SD algorithm has been considered in several works \cite{sd9,sd10,Pauli}. The computational complexity of the SD algorithm for  uncoded systems is analyzed in \cite{sd9,sd10}. On the other hand,\cite{Pauli} analyzes the SD complexity of a multiple symbol differentially encoded multiple-input multiple-output (MIMO) system. For an encoded MIMO system, increasing the diversity comes at the price of decreasing the spatial multiplexing gain. An appropriate criterion for the achievable performance in multiple antenna systems is the diversity multiplexing tradeoff (DMT) \cite{tsebook}. Also, \cite{jal11} incorporates the space-time codes into the analysis and introduces the concept of complexity exponent as a functions of multiplexing gain, $r$, denoted by $c(r)$. Indeed, $\rho^{c(r)}$, where $\rho$ is the signal to noise ratio (SNR), determines the minimum necessary complexity for achieving the DMT optimal performance using a conventional SD algorithm. The analytical results  in \cite{jal11} show that $c(r)\neq 0$ for $0<r<min(n_r,n_t)$ which means that the complexity of conventional SD algorithms increases polynomially with $\rho$. In other words, the computational complexity of a conventional DMT optimal SD algorithm tends to infinity at high SNR. 

In this paper, we propose a DMT optimal pruning for the SD algorithm, which has a constant average complexity at high SNR. More precisely, the proposed SD algorithm visits a single branch on average, i.e. one node per layer, at high SNR. The main contributions of our paper are listed as follows.
\begin{itemize}
	\item By intersecting some one dimensional spheres, here after referred to as zero-spheres\footnote{The $n$-sphere is a generalized sphere  an $n+1$ dimensional Euclidean space and zero-sphere represents the boundary of a line segment.}, with the well-known hypersphere of the SD algorithm, we propose a pruning method. The zero-spheres are constructed around the minimum mean squared error (MMSE) equalized observations which trap some of constellation
	points within a zero-sphere with radius $R_{\rm ZSD}$. These selected constellation points are intersected with the lattice points that lie inside the well-known hypersphere of radius $R_{\rm SD}$. The radii $R_{\rm ZSD}$ and $R_{\rm SD}$ are proposed such that the pruning
	brings about a DMT optimal decoding.
	\item We show that unlike the computational complexity of the conventional DMT optimal SD algorithm analyzed in \cite{jal11}, which increases with $\rho$, the complexity of the proposed algorithm {\it is not} an increasing function of $\rho$, and visiting a single node per layer is
	sufficient to achieve DMT optimality.
	\end{itemize}
The rest of the paper is organized as follows. Section \ref{systemmodel11} introduces the system model of a layered space-time block code (LSTBC)  MIMO system, which we select as a reference model although the applicability of our study is not limited by LSTBC MIMO systems.  The proposed DMT optimal pruning algorithm is presented in Section \ref{proposed algorithm}. Simulation results are reported and discussed in Section \ref{simres}, and Section \ref{conclusion} summarizes the paper.

{\it Notation:} We denote the vectors by
lowercase bold letters $\bx$, and matrices by uppercase bold letters $\bX$ with $[\bX]_{i,j}$ being the $ij$th element. We let $\mathds{R}$ and $ \mathds{C}$ to denote the set of real and complex numbers, respectively. Also $\varnothing$ denotes the empty set. The transpose of a vector $\bx$ is $\bx^{T}$ and $\|\bx\|$ is the Euclidean norm of $\bx$. We utilize $\textswab{Re}\{x\}$ and  $\textswab{Im}\{x\}$ for real and imaginary parts of $x$. The symbols $\doteq$, $\dot{\leq}$, and $\dot{\geq}$ denote the asymptotic exponential equality and inequalities, respectively, i.e., $g(\rho) \dot{=} h(\rho)$ means $\lim \limits_{\rho\rightarrow \infty}\frac{\log g(\rho)}{\log \rho}=\lim \limits_{\rho\rightarrow \infty}\frac{\log h(\rho)}{\log \rho}$ and the definition of $\dot{\leq}$ and $\dot{\geq}$ can be obtained by replacing $=$ with $\leq$ and $\geq$ in the given expression, respectively. 
\section{System Model} \label{systemmodel11}

Consider a MIMO communication system with $n_t$ transmit and $n_r$ receive antennas. The input-output baseband relationship with a signal fading block of $ T$ symbol durations, can be expressed in the form
\beq \label{systemmodelraw}
\bY = \bH_c \bX + \bW,
\eeq
where   $\bY \in  \mathds{C}^{n_r\times  T} $ is the received signal matrix and $T$ is the coherence time. The $[\bH_c]_{i,j}$ element of the channel matrix $\bH_c \in \mathds{C}^{n_r\times n_t}$ corresponds to the channel coefficient between the $j$th transmit and the $i$th receive antennas and it is considered to be  independent identically distributed (i.i.d) circularly symmetric complex Gaussian zero mean random variable of unit variance. Furthermore, the matrix  $\bW \in  \mathds{C}^{n_r\times  T} $ is the spatially and temporally additive noise with i.i.d circularly symmetric complex Gaussian components with zero mean and variance $\sigma^2$. Finally, $\bX$ is the $n_t\times  T$ codeword matrix, whose $i,k$th element corresponds  to the signal transmitted from the $i$th antenna during the $k$th  symbol interval for $1\leq k\leq  T$.

For the decoding process, we vectorize the observation matrix $\bY$ as,
$
\by = {\vec (\bY)}= \left[\textswab{Re}\{[\bY]_{1,1}\},\textswab{Im}\{[\bY]_{1,1}\},\dots,\textswab{Re}\{[\bY]_{n_r,1}\},\textswab{Im}[\bY]_{n_r,1}\},  \dots,\right. \\ \left. \textswab{Re}\{[\bY]_{1,T}\}, \textswab{Im}[\bY]_{1,T}\},  \dots,\textswab{Re}\{[\bY]_{n_r,T}\},\textswab{Im}\{[\bY]_{n_r,T}\} \right]^\rT$, where $\by \in \mathds{R}^{2n_rT}$. By defining  the complex to real mapping operator $r_i(\bH_c)$, which maps each complex element of $\bH_c$ as
 \beq
  \left(\begin{array}{cc}
 \textswab{Re}\{[\bH]_{i,j}\}&-\textswab{Im}\{[\bH]_{i,j}\}\\
 \textswab{Im}\{[\bH]_{i,j}\}&\textswab{Re}\{[\bH]_{i,j}\}
 \end{array}\right),
 \eeq
 the system model (\ref{systemmodelraw}) can be rewritten as 
 \beq \label{first_model}
 \by = \mathcal{H} \bx + \bw,
 \eeq
 where $\bx = \vec{(\bX)}$, $\bw = \vec{(\bW)}$, and $\mathcal{H}=\diag (r_i(\bH_c),\dots,r_i(\bH_c))\in \mathds{R}^{2n_rT\times2n_tT}$ is a block diagonal matrix with $r_i(\bH_c)$ on its main diagonal \cite{sd3}.
 
 For an LSTBC, we have $\bx = \bC \bs$, where $\bC \in \mathds{R}^{2n_tT\times2n_tT}$ is the encoding matrix and $\bs \in \mathcal{C}^{2n_tT}$. We assume that $\mathcal{C}$ is a pulse amplitude modulated (PAM) signal set of size $Q$, i.e.,
 \beq
 \mathcal{C} = \{c_i = (2i - Q + 1)d_Q| i \in \mathds{Z}_Q\},
 \eeq 
 with $\mathds{Z}_Q \triangleq \{0, 1,\dots, Q-1\}$, where $Q = |\mathcal{C}|$ is the cardinality of the set $\mathcal{C}$ and $d_Q$ is the minimum distance between the constellation points. 
 Hence, the system model in (\ref{first_model}) can be rewritten as
 \beq \label{systemmodel}
 \by = \bH \bs + \bw
 \eeq
 where $\bH = \mathcal{H}\bC \in \mathds{R}^{2n_rT\times2n_tT}$ \cite{sd3}. 
 
 Let $\mathcal{R}(\rho_T)$ be the normalized data rate. then spatial multiplexing gain can obtained as\cite{tsebook}
 \beq
 r=\lim_{\rho\rightarrow\infty}\frac{\mathcal{R}(\rho_T)}{\log\rho_T},
 \eeq
 and the DMT of a decoding scheme, i.e., $d(r)$, is given by
 \beq\label{3322}
 d(r)=-\lim_{\rho_T\rightarrow\infty}\frac{\log\rP_\re(\rho_T)}{\log(\rho_T)},
 \eeq
 where the SNR $\rho_T\triangleq n_t\rho$, $\rho=\frac{1}{\sigma^2}$ and $\rP_\re(\rho_T)$ is the error probability of the decoding scheme. 
 \section{The Proposed DMT optimal Pruning algorithm}\label{proposed algorithm}
 In this section, we present a DMT optimal pruning algorithm for the decoding problem  (\ref{systemmodelraw}). Besides being DMT optimal, we show that the proposed algorithm benefits from an appropriate complexity behavior at high SNR. More precisely, we show that, as the SNR increases and for high data transmission rates, the computational complexity of the proposed algorithm becomes constant, and visiting only a single node is sufficient to obtain DMT optimality.  Two types of tree pruning are considered in the $k$th layer of the proposed algorithm. This pruning is the result of the intersection of a conventional $k+1$ dimensional hypersphere and a one dimensional zero-sphere with different centers and
 radii. Our aim is to design the radii of these two spheres in order to guarantee the DMT optimality.  If necessary, the reader is encouraged to take a look at \cite{sd9} before proceeding with the rest of the paper, in order to have a thorough understanding of the conventional SD algorithms.
 
 In the proposed method, the zero-sphere is formed based on the MMSE equalized observation, i.e.,
 \beq
 \tilde{\by} = (\bH^T \bH+\frac{1}{\rho}\bI)^{-1} \bH^T \by.
 \eeq
 Indeed, $2n_tT$ zero-spheres of radius $R_{\rm ZSD}^{(k)}$, centered at $\tilde{y}_k$ are formed as
 \beq \label{eq9}
 |s_k-\tilde{y}_k|^2<\left(R_{{\rm ZSD}}^{(k)}\right)^2,
 \eeq
 where $\tilde{y}_k$ is the $k$th equalized observation. The constellation points that lie inside the zero-spheres are obtained as 
 \beq \label{szsd}
\mathcal{S}_{{\rm ZSD}}^{(k)}=\{c_i\in\mathcal{C}|c_i\in\mathfrak{I}_{{\rm ZSD}}^{(k)},i\in\mathbb{Z}_Q\},
\eeq
 where the interval $\mathfrak{I}_{{\rm ZSD}}$ is 
\beq
\mathfrak{I}_{{\rm ZSD}}^{(k)}=[s_{{\rm min,ZSD}}^{(k)},s_{{\rm max,ZSD}}^{(k)}],
\eeq
and 
\beq
s_{{\rm min,ZSD}}^{(k)}=\min\{c_{Q-1},\tilde{y}_k-R_{{\rm ZSD}}^{(k)}\},
\eeq
\beq
s_{{\rm max,ZSD}}^{(k)}=\max\{c_{0},\tilde{y}_k+R_{{\rm ZSD}}^{(k)}\}.
\eeq
 In accordance with the conventional SD algorithms, the $k+1$ dimensional hyperspheres generate some sets which are specified
 as\cite{sd9}
 \beq \label{ssd}
 \mathcal{S}_{{\rm SD}}^{(k)}=\{c_i\in\mathcal{C}|c_i\in\mathfrak{I}_{{\rm SD}}^{(k)},i\in\mathbb{Z}_Q\},
 \eeq
 where
 \beq
 \mathfrak{I}_{{\rm SD}}^{(k)}=[s_{{\rm min,SD}}^{(k)},s_{{\rm max,SD}}^{(k)}],
 \eeq
 and
  
 \beq
 s_{{\rm min,SD}}^{(k)}=\min\left\{c_{Q-1},\frac{-R_k+\bar{y}_{k|k+1}}{[\bR]_{k,k}}\right\},
 \eeq
 \beq
 s_{{\rm max,SD}}^{(k)}=\max\left\{c_{0},\frac{R_k+\bar{y}_{k|k+1}}{[\bR]_{k,k}}\right\},
 \eeq
 where $\bar{y}_{k|k+1}=\bar{y}_k-\sum_{j=k+1}^{2n_tT}[\bR]_{k,j}s_j$ and $R_{k}^2=R_{k+1}^2-(\bar{y}_{k+1|k+2}-[\bR]_{k+1,k+1}s_{k+1})^2$ and we have $R_{2n_tT}^2=R_{\rm SD}^2-\|\bQ_2^{H}\by\|^2$ and $\bar{\by}=\bQ_1^H\by$. Moreover, $\bQ$ and $\bR$ are obtained from QR factorization of $\bH$ as
 \beq
 \bH=[\bQ_1\; \bQ_2]\begin{bmatrix} \bR \\ \mathbf{0}_{(n_r-n_t)\times n_t} \end{bmatrix},
 \eeq
 where $\bQ_1$ and $\bQ_2$ them matrices of the first $n_t$ and remaining $n_r-n_t$ orthonormal columns of $\bQ$, respectively, and $\bR$ is the $n_t\times n_t$ upper triangular matrix with positive diagonal elements.
 
 The intersection of $\mathcal{S}^{(k)}_{\rm ZSD}$ and $\mathcal{S}^{(k)}_{\rm SD}$ defined in (\ref{szsd}) and (\ref{ssd}) at the $k$th layer of the algorithm, results in a search tree which is jointly pruned via the zero-spheres and the hypersphere of the SD algorithm. 
 
  The proposed algorithm is summarized as Algorithm 1.

	\SetAlgoNlRelativeSize{0}
	
	\begin{algorithm}
			\SetNlSty{textbf}{}{.}
		\KwIn{$\bQ=[\bQ_1 \bQ_2]$, $\bR$, ${\by}$, $R_{\rm ZSD}$, $R_{\rm SD}$}
		\KwOut{$\hat{\bs}$}
		  Set $k=2n_tT$, $R_{2n_tT}^2=R_{\rm SD}^2-\|\bQ_2^{H}\by\|^2$, $\bar{y}_{2n_tT|2n_tT}=\bar{y}_{2n_tT}$, $i=-1$.\\
		Set $\mathcal{S}^{(k)}=\mathcal{S}_{{\rm ZSD}}^{(k)}\cap\mathcal{S}_{{\rm SD}}^{(k)}$, if $\mathcal{S}_{{\rm ZSD}}^{(k)}\cap\mathcal{S}_{{\rm SD}}^{(k)}=\varnothing$, $\mathcal{S}^{(k)}=\mathcal{S}_{{\rm ZSD}}^{(k)}$.\\
		$i=i+1$, if $c_i\in \mathcal{S}_k$, $s_k=c_i$, go to step 5, else go to step 4.\\
		$k=k+1$ if $k=2n_tT+1$ terminate algorithm, else go to step 3.\\
	 if $k=1$ go to step 6. Else $k=k-1$, $\bar{y}_{k|k+1}=\bar{y}_k-\sum_{j=k+1}^{2n_tT}[\bR]_{k,j}s_j$, $R_k^2=R_{k+1}^2-(\bar{y}_{k+1|k+2}-[\bR]_{k+1,k+1}s_{k+1})^2$ and go to 2.\\
	  $\hat{\bs}$ is found. Save $\bs$ and its distance from $\by$, $R_{2n_tT}^2=R_1^2+(\bar{y}_1-[\bR]_{1,1,}s_1)^2$, go to step 3.
	\caption{DMT-optimal Tree Pruning SD}
	\label{algorithm}
	\end{algorithm}
Now, our aim is to determine $\mathcal{S}_{{\rm SD}}^{(k)}$ and $\mathcal{S}_{{\rm ZSD}}^{(k)}$ such that the DMT optimality is guaranteed. To this end the redii corresponding to these sets are obtained.

{\bf Definition 1.} Let the error performance gap between the proposed method and the maximum likelihood (ML) decoding, i.e. $\rP_{\rm ML} - \rP_{\rm e}$, where $\rP_{\rm ML}$ and
$\rP_{\rm e}$ are the error probabilities of the ML decoding and the proposed method. We denote the upper bound for this
performance gap by $\Delta \rP$.

{\bf Theorem 1.} 
	Choosing $R_{\rm ZSD}^{(k)}=d_Q+\frac{\left(d_{\rm ML}(r) -d_{\rm MMSE}(r)\right)\ln\rho}{d_Q\rho_{{\rm MMSE}}^{(k)}}$ and $R_{\rm SD}^2=d_{\rm ML}(r){\frac{\ln \rho}{\rho}}$ leads to $\Delta \rP \doteq \rho^{-d_{ML}(r)}$ where  $d_{\rm MMSE}(r)$ and $d_{\rm ML}(r)$ are the DMT of MMSE and ML methods and $\rho_{{\rm MMSE}}^{(k)}=\frac{\rho}{[(\bH^\rT\bH + \rho^{-1}\bI)^{-1}]_{kk}}- 1$.
\begin{proof}
	Defining $\mathcal{S}_{\rm ZSD}\cap\mathcal{S}_{\rm SD}=\mathcal{S}$, one can write
	\begin{align}\label{1214}
	&\rP(\hat{\bs}\neq \bs|\bH)=\nonumber\\&\rP(\hat{\bs}\neq \bs|\bs\in\mathcal{S},\bH) \rP(\bs\in\mathcal{S}|\bH)+\nonumber\\&\rP(\hat{\bs}\neq \bs|\bs\notin\mathcal{S},\bH) \rP(\bs\notin\mathcal{S}|\bH).
	\end{align}
	The reduce search space is also $\mathcal{S} =\prod_{i=1}^K  \mathcal{S}^{(i)}$ which is $K$ary Cartesian product over $K$ set $\{\mathcal{S}^{(i)}\}_{i=1}^K$. It is readily seen that
	\beq
	\rP(\hat{\bs}\neq\bs|\bs\in\mathcal{S},\bH)\leq\rP(\hat{\bs}_{\rm ML}\neq\bs|\bH).
	\eeq
	On the other hand, since $\rP(\hat{\bs}\neq\bs|\bs\notin\mathcal{S},\bH)\leq1$, we have
	\beq\label{1810}
	\rP(\hat{\bs}\neq\bs|\bH)\leq\rP(\hat{\bs}_{\rm ML}\neq\bs|\bH)+\rP(\bs\notin\mathcal{S}|\bH).
	\eeq
	Using union bound, the second term of the above inequality can be written as 
	\beq\label{1811}
	\rP(\bs\notin\mathcal{S}|\bH)=\rP(\bigcup_{k=1}^{2n_tT}s_k\notin\mathcal{S}^{(k)}|\bH)\leq\sum_{k=1}^{2n_tT}\rP(s_k\notin\mathcal{S}^{(k)}|\bH).
	\eeq
	Now we expand $\rP(s_k\notin\mathcal{S}^{(k)}|\bH)$ as follows
	\begin{align}\label{1813}
	&\rP(s_k\notin\mathcal{S}^{(k)}|\bH)=\nonumber\\&\rP(s_k\notin\mathcal{S}^{(k)}|\bH,\mathcal{S}_{\rm ZSD}^{(k)}\cap\mathcal{S}_{\rm SD}^{(k)}=\varnothing)\rP(\mathcal{S}_{\rm ZSD}^{(k)}\cap\mathcal{S}_{\rm SD}^{(k)}=\varnothing|\bH)+\nonumber\\&\rP(s_k\notin\mathcal{S}^{(k)}|\bH,\mathcal{S}_{\rm ZSD}^{(k)}\cap\mathcal{S}_{\rm SD}^{(k)}\neq\varnothing)\rP(\mathcal{S}_{\rm ZSD}^{(k)}\cap\mathcal{S}_{\rm SD}^{(k)}\neq\varnothing|\bH)\nonumber\\&=\rP(s_k\notin\mathcal{S}_{\rm ZSD}^{(k)}|\bH)\rP(\mathcal{S}_{\rm ZSD}^{(k)}\cap\mathcal{S}_{\rm SD}^{(k)}=\varnothing|\bH) \nonumber\\&+\rP(s_k\notin\mathcal{S}_{\rm ZSD}^{(k)}\cap\mathcal{S}_{\rm SD}^{(k)}|\bH)\rP(\mathcal{S}_{\rm ZSD}^{(k)}\cap\mathcal{S}_{\rm SD}^{(k)}\neq\varnothing|\bH).
	\end{align}
	It should be noted that in the proposed algorithm, when $\mathcal{S}_{\rm ZSD}^{(k)}\cap\mathcal{S}_{\rm SD}^{(k)}=\varnothing$ we have $\mathcal{S}^{(k)}=\mathcal{S}^{(k)}_{\rm ZSD}$.  Using the union bound, one can show that
	\begin{align}\label{1812}
	\rP(s_k\notin\mathcal{S}_{\rm ZSD}^{(k)}\cap\mathcal{S}_{\rm SD}^{(k)}|\bH)\leq\rP(s_k\notin\mathcal{S}_{\rm ZSD}^{(k)}|\bH)+\rP(s_k\notin\mathcal{S}_{\rm SD}^{(k)}|\bH).
	\end{align}
	Therefore, using (\ref{1813}) and (\ref{1812}), we have 
	\begin{align}\label{1814}
	\rP(s_k\notin\mathcal{S}^{(k)}|\bH)\leq\rP(s_k\notin\mathcal{S}_{\rm ZSD}^{(k)}|\bH)+\rP(s_k\notin\mathcal{S}_{\rm SD}^{(k)}|\bH).
	\end{align}
	Consequently, employing (\ref{1810}), (\ref{1811}), (\ref{1814}) and the expansion of  $\rP_\re$ as
	\beq
	\rP_{\rm e} = \rE_{\bH} \left\{ \rE_{\bs}\left\{{\rm{P}}_{\bs}(\hat{\bs}\neq \bs|\bH)\right\} \right\}, 
	\eeq
	and considering that $\rP_{\rm e}\geq\rP_{\rm e}^{\rm ML}$, we have 
	\beq \label{deltap}
	\rP_{\rm e}^{\rm ML}\leq \rP_{\rm e} \leq \rP_{\rm e}^{\rm ML} + \Delta \rP .
	\eeq
	where $\Delta \rP=\Delta \rP_{\rm ZSD} + \Delta \rP_{\rm SD}$ and $\rP_{\rm e}^{\rm ML}=\rE_{\bH,\bs}\left\{ {\rm{P}}_{\bs}(\hat{\bs}_{\rm ML}\neq \bs|\bH)\right\}$ is the symbol error probability of the ML detector and 
	\beq\label{1218}
	\Delta \rP_{\rm ZSD} = \rE_{\bH}\left\{\sum_{k=1 }^{2n_tT}\left\{\rP_\bs{(\bs \notin \mathcal{S}_{\rm ZSD}|\bH)}\right\}\right\},
	\eeq
	\beq\label{1229}
	\Delta \rP_{\rm SD} = \rE_{\bH}\left\{\sum_{k=1 }^{2n_tT}\{\rP_\bs{(\bs \notin \mathcal{S}_{\rm SD}|\bH)}\}\right\}.
	\eeq
	Note that, $\rP(A) = \rE_{s_k}\left\{\rP_{s_k}(A)\right\}$ where $\rP_{s_k}(A) =\rP(A|s_k)$. Now, we focus on calculating $\Delta \rP_{\rm ZSD}$
	\begin{align}\label{1219}
	\rP_{\bs}(\bs \notin \mathcal{S}_{\rm ZSD}|\bH)=\sum_{k=1}^{2n_tT}\rP_{s_k}(s_k\notin\mathcal{S}_{{\rm ZSD}}^{(k)}|\bH).
	\end{align}
	Considering $s_k=c_i=(2i-Q+1)d_Q$, we can expand the above equation over the events $c_i\in \mathcal {I}$ and $c_i\in \mathcal {O}$, where $\mathcal{I}$ and $\mathcal{O}$ are the inner and the outer modulation points set. Consequently,
	\begin{align}\label{1211}
	&\rP_{s_k}(s_k \notin \mathcal{S}_{{\rm ZSD}}^{(k)}|\bH)=\nonumber\\&\rP_{s_k}(s_k \notin \mathcal{S}_{{\rm ZSD}}^{(k)}|\bH,s_k\in \mathcal{I})\rP(s_k\in\mathcal{I}|\bH)+\nonumber\\& \rP_{s_k}(s_k \notin \mathcal{S}_{\rm ZSD}^{(k)}|\bH,s_k\in \mathcal{O})\rP(s_k\in\mathcal{O}|\bH),
	\end{align}
	where $\rP(s_k\in\mathcal{I}|\bH)=\frac{Q-2}{Q}$ and $\rP(s_k\in\mathcal{O}|\bH)=\frac{2}{Q}$.
	We have
	\begin{align}\label{1220}
	&\rP_{s_k}(s_k \notin \mathcal{S}_{{\rm ZSD}}^{(k)}|\bH,s_k\in \mathcal{I})=\nonumber\\&1- \rP_{s_k}(s_{{\rm min,ZSD}}^{(k)}\leq c_i \leq s_{{\rm max,ZSD}}^{(k)}|\bH).
	\end{align} 
	Using (\ref{eq9}) we have
	\begin{align}
	&\rP_{s_k}(s_{{\rm min,ZSD}}^{(k)}\leq c_i \leq s_{{\rm max,ZSD}}^{(k)}|\bH)=\nonumber\\& \rP_{s_k}(c_i-R_{{\rm ZSD}}^{(k)}\leq\tilde{y}_k\leq c_i+R_{{\rm ZSD}}^{(k)}|\bH)
	\end{align}
	According to $\tilde{y}_k=c_i+\tilde{w}_k$ and the above equations, one can show that
	\begin{align}\label{1221}
	&\rP_{s_k}(s_k \notin \mathcal{S}_{{\rm ZSD}}^{(k)}|\bH,s_k\in \mathcal{I})=\nonumber\\&1-\rP_{s_k}(-R_{{\rm ZSD}}^{(k)}\leq\tilde{w}_k\leq R_{{\rm ZSD}}^{(k)}|\bH)=\nonumber\\&2\mathcal{Q}\left(R_{{\rm ZSD}}^{(k)}\sqrt{\rho_{{\rm MMSE}}^{(k)}}\right),
	\end{align}
	where  $\rho_{{\rm MMSE}}^{(k)}\triangleq\rho z_k$ \cite{zeroforcing}
	\beq\label{9091}
	z_k  = \frac{1}{[(\bH^\rT\bH + \rho^{-1}\bI)^{-1}]_{kk}}- \frac{1}{\rho}.
	\eeq
	Similarly for outer constelation points we have,
	\begin{align}\label{1222}
	&\rP_{s_k}(s_k\notin\mathcal{S}_{{\rm ZSD}}^{(k)}|\bH,s_k\in\mathcal{O})=\rP(s_{{\rm min,ZSD}}^{(k)}>c_0)\nonumber\\&=\rP(s_{{\rm max,ZSD}}^{(k)}<c_{Q-1})=\mathcal{Q}(R_{{\rm ZSD}}^{(k)}\sqrt{\rho_{{\rm MMSE}}^{(k)}}).
	\end{align}
	Substituting (\ref{1221}) and (\ref{1222}) in (\ref{1211}), we conclude that
	\begin{align}\label{1223}
	\sum_{k=1}^{2n_tT}\rP_{s_k}(s_k\notin\mathcal{S}_{{\rm ZSD}}^{(k)}|\bH)=\frac{2(Q-1)}{Q}\sum_{k=1}^{2n_tT}\mathcal{Q}\left(R_{{\rm ZSD}}^{(k)}\sqrt{\rho_{{\rm MMSE}}^{(k)}}\right).
	\end{align}
	As it was mentioned, our choice of $R_{\rm ZSD}^{(k)}$ is $R_{\rm ZSD}^{(k)}=d_Q+\Delta R^{(k)}$. It should be noted that if $\Delta R^{(k)}=0$, the MMSE point merely falls inside the sphere. Obviously, $\Delta R^{(k)}=0$ only achieves the DMT of MMSE and does not arrive at the DMT optimality. Thus, in order to avoid this, we consider a chance for the other points to lie inside the zero sphere. The added points should be selected such that it gaurantees the DMT optimality. If we choose $\Delta R^{(k)}$ as a multiple of $d_Q$, i.e. $\Delta R^{(k)}=\alpha d_Q$, it is obvious that increasing $\alpha$ leads to more additional points inside the sphere.
	
	A reasonable choice is selecting $\alpha$ such that it decreases with effective SNR of the MMSE method. i.e. $\frac{d_Q^2}{(\sigma_{{\rm MMSE}}^{(k)})^2}$, where $(\sigma_{{\rm MMSE}}^{(k)})^2=\frac{1}{\rho_{{\rm MMSE}}^{(k)}}$. Therefore, a hurestic choice of $\alpha$ could be $\alpha=\gamma(\frac{\sigma_{{\rm MMSE}}^{(k)}}{d_Q})^2$. Thus, our choice for $\Delta R^{(k)}$ is 
	\beq
	\Delta R^{(k)}=\frac{\gamma}{d_Q\rho_{{\rm MMSE}}^{(k)}}.
	\eeq
	In the sequel, we determine $\gamma$ such that it gaurantees the DMT optimality of the proposed detector. According to $R_{{\rm ZSD}}^{(k)}=d_Q+\frac{\gamma}{d_Q\rho_{{\rm MMSE}}^{(k)}}$ and the above discussion and (\ref{1223}) we have
	
	\beq \label{eq39}
	\Delta \rP_{\rm ZSD} = \frac{2(Q-1)}{Q} \sum_{k=1}^{2n_tT}  \rE_{Z_k} \left\{ \mathcal{Q}\left(\frac{\gamma}{d_Q\sqrt{ \rho  z_k}}+{d_Q\sqrt{ \rho z_k} }\right)\right\}.
	\eeq
	In order to calculate $\Delta P_{\rm ZSD}$ we need the following lemma.
	
	{\bf Lemma 1:} By  defining $x_k = \left([(\bH^\rT \bH)^{-1}]_{kk} \right)^{-1}$, for a positive $g(z_k)$ where $z_k$ is defined as (\ref{9091})
	\beq
	\int g(z_k)f_{Z_k}(z_k)dz_k \leq \int g(x_k) f_{X_k}(x_k) dx_k.
	\eeq
	\begin{proof} 
		See the Appendix A. 
	\end{proof}
	According to Lemma 1 we have 
	\begin{align}\label{98981}
	&\rE_{Z_k} \left\{ \mathcal{Q}\left(\frac{\gamma}{d_Q\sqrt{ \rho  z_k}}+{d_Q\sqrt{ \rho z_k} }\right)\right\} \nonumber\\
	&\leq \rE_{X_k} \left\{ \mathcal{Q}\left(\frac{\gamma}{d_Q\sqrt{ \rho  x_k}}+{d_Q\sqrt{ \rho x_k} }\right)\right\}.
	\end{align}
	Since, $[(\bH^\rT\bH)^{-1}]_{kk} = \be_k^\rT (\bH^\rT\bH)^{-1} \be_k$ where $\be_k = [0,\dots,0,1,0,\dots,0]^\rT$ is a vector with  zero elements except for it's $k$th element, we have
	\beq
	[(\bH^\rT\bH)^{-1}]_{kk} = \be_k^\rT \bR^{-1}(\mathcal{H}^\rT\mathcal{H})^{-1} (\bR^\rT)^{-1} \be_k.
	\eeq
	Assuming $\bl_k \triangleq(\bR^\rT)^{-1}\be_k$ and $\bl_k = [\bl_k(1),\dots,\bl_k(T)]^\rT$ we have
	\beq
	x_k = \frac{1}{\sum_{i=1}^T \bl_k^\rT(i)(r_i(\bH_c)^\rT r_i(\bH_c))^{-1}\bl_k(i)}.
	\eeq
	Assuming $\arg \max \limits_i \bl_k^\rT(i)(r_i(\bH_c)^\rT r_i(\bH_c))^{-1}\bl_k(i)=m$, one can show that
	\beq \label{namosavi}
	{x_k} \leq {t_k},
	\eeq
	where $t_k = \frac{1}{\bl_k^\rT(m)(r_m(\bH_c)^\rT r_m(\bH_c))^{-1}\bl_k(m)}$. In \cite{jerry}, it is shown that the variable $\frac{\| \bl_k^\rT(m)\|^2}{\bl_k^T(m)(r_m(\bH_c)^\rT r_m(\bH_c))^{-1}\bl_k(m)}$ is chi-squared distributed as $\mathcal{X}^2_{2(n_r - n_t +1)}$.
	
	Now we calculate the following expected value in (\ref{98981}) as
	\begin{align}
	&\rE_{X_k} \left\{ \mathcal{Q}\left(\frac{\gamma}{d_Q\sqrt{ \rho  x_k}}+{d_Q\sqrt{ \rho x_k} }\right)\right\} \nonumber \\
	&= \frac{1}{\rho} \int_0^\infty  \mathcal{Q}\left(\frac{\gamma}{d_Q\sqrt{   x_k}}+{d_Q\sqrt{  x_k} }\right)f_{X_k}(\frac{x_k}{\rho}) dx_k.
	\end{align}
	Invoking the fact that $\lim_{dx_k\rightarrow 0} f_{X_k}(\frac{x_k}{\rho})dx_k =\lim_{dx_k\rightarrow 0} \rP(\frac{x_k}{\rho}\leq X_k\leq \frac{x_k}{\rho}+dx_k)$ and the inequality (\ref{namosavi}), results in
	\beq
	\lim_{dx_k\rightarrow 0}\rP(\frac{x_k}{\rho}\leq X_k\leq \frac{x_k}{\rho}+dx_k) \leq \lim_{dx_k\rightarrow 0}\rP(0\leq T_k\leq \frac{x_k}{\rho}+dx_k),
	\eeq
	and also
	\beq
	\lim_{\rho\rightarrow \infty,{dx_k\rightarrow 0}} \rP(0\leq T_k\leq \frac{x_k}{\rho}+dx_k)  = f_{T_k}(\frac{x_k}{\rho})dx_k, 
	\eeq
	therefore
	\begin{align}
	&\rE_{X_k} \left\{ \mathcal{Q}\left(\frac{\gamma}{d_Q\sqrt{ \rho  x_k}}+{d_Q\sqrt{ \rho x_k} }\right)\right\} \nonumber \\
	&\leq\int_0^\infty \mathcal{Q}\left(\frac{\gamma}{d_Q\sqrt{   \rho x_k}}+{d_Q\sqrt{ \rho x_k} }\right) f_{T_k}(x_k) dx_k.
	\end{align}
	According to the inequality $\mathcal{Q}(x)\leq \frac12 e^{-\frac{x^2}{2}}$ we have
	\begin{align} \label{222}
	&\rE_{X_k} \left\{ \mathcal{Q}\left(\frac{\gamma}{d_Q\sqrt{ \rho  x_k}}+{d_Q\sqrt{ \rho x_k} }\right)\right\} \nonumber \\
	&\leq\frac{e^{-\gamma}}{2}\int_0^\infty e^{-\rho h(x_k)} f_{T_k}(x_k) dx_k,
	\end{align}
	where $h(x_k) = \frac12 \left(\frac{\gamma^2}{d^2_Q\rho^2x_k}+d_Q^2 x_k \right) $. Using the Laplace method \cite{1994lap} we have 
	\beq \label{laplace}
	\lim \limits_{\rho\rightarrow \infty} \int_a^b f(t)e^{-\rho h(t)}dt\approx \sqrt{\frac{2\pi}{\rho|h^{\prime \prime}(t_0)|}}f(t_0) e^{-\rho h(t_0)},
	\eeq
	where $h^{\prime\prime}(t_0)$ is the second derivative of $h(t)$ and $t_0$ is the root of the derivative of $h(t)$. Applying (\ref{laplace}) to the right hand side of  (\ref{222}) and since $t_0= \frac{\gamma}{d_Q^2\rho}$, yields
	\begin{align}  \label{eq51}
	&\rE_{X_k} \left\{ \mathcal{Q}\left(\frac{\gamma}{d_Q\sqrt{ \rho  x_k}}+{d_Q\sqrt{ \rho x_k} }\right)\right\}\nonumber\\ &\leq \sqrt{\frac{2\pi \gamma}{\rho^2 d_Q^4}}e^{-2\gamma} f_{T_k}\left(\frac{\gamma}{d_Q^2\rho}\right).
	\end{align}
	As it was previously stated 
	\beq\frac{\| \bl_k^\rT(m)\|^2}{\bl_k^\rT(m)(r_i(\bH_c)^\rT r_i(\bH_c))^{-1}\bl_k(m)}\sim \mathcal{X}^2_{2(n_r - n_t +1)},\eeq
	therefore
	\beq \label{eq53}
	f_{T_k}\left( \frac{\gamma}{d_Q^2\rho}\right) \doteq \left(\frac{\gamma}{d_Q^2\rho} \right)^{n_r  - n_t}.
	\eeq
	Thus, according to (\ref{eq39}), (\ref{98981}), (\ref{eq51}) and (\ref{eq53}) we have
	\beq
	\Delta P_{\rm ZSD} \dot{\leq} \frac{1}{\sqrt{2\gamma}} e^{-2\gamma}(\frac{2\gamma}{d_Q^2 \rho})^{n_r-n_t+1}.
	\eeq
	Since the average energy of the transmitting symbols in quadrature amplitude modulation (QAM) is $\rE_{\rm QAM} = \frac{d_Q^2(2^{\mathcal{R}(\rho_T)}-1)}{6}$, for a normalized modulation scheme, i.e. $\rE_{\rm QAM}  = 1$, we have $d_Q^2 =\frac{6}{2^{\mathcal{R}(\rho_T)}-1}$. Therefore, according to $\mathcal{R}(\rho_T) \doteq \frac{r}{n_t}\log \rho$ we get
	\beq
	d^2_Q \doteq \rho^{-\frac{r}{n_t}},
	\eeq  
	and consequently
	\beq \label{eq56}
	\Delta P_{\rm ZSD} \dot{\leq} \gamma^{n_r-n_t+\frac12}e^{-2\gamma}\rho^{-(n_r-n_t+1)(1-\frac{r}{n_t})}.
	\eeq
	
	According to (\ref{eq56}), in order for the proposed method to achieve the optimum DMT we can choose
	\beq
	\gamma = \psi(r)\ln \rho,
	\eeq 
	where 
	\beq
	\psi(r)  = d_{\rm ML}(r) -d_{\rm MMSE}(r).
	\eeq
	In other words, based on (\ref{deltap}), with this choice of $\gamma$ we have $\rP_\re \dot{=}\rP_{\rm ML}$ where $d_{\rm MMSE}(r)=(n_r-n_t+1)(1-\frac{r}{n_t})$. 
	
	On the other hand, $\Delta \rP_{\rm SD}$ can be written as,
	\beq \label{eq59}
	\Delta \rP_{\rm SD} =\rE_{\bH}\left\{ \rE_{\bs_k}\left\{\sum_{k=1}^{2n_tT}\rP_{\bs_k}(s_k\notin\mathcal{S}_{{\rm SD}}^{(k)}|\bH)\right\}\right\}.
	\eeq
	where 
	\begin{align}\label{12111}
	\rP_{\bs_k}(s_k\notin\mathcal{S}_{{\rm SD}}^{(k)}|\bH)\leq\rP_{\bs}(\bs \notin \mathcal{S}_{\rm SD}|\bH)
	\end{align}
	which leads to
	\begin{align}\label{12113}
	\rP_{\bs}(\bs \notin \mathcal{S}_{\rm SD}|\bH)=\rP(\|\bw\|>R_{\rm SD}),
	\end{align}
	where $\|\bw\|^2$ is the centeralized chi-squared random variable with $4n_rT$ degrees of freedom. Therefore,
	\begin{align}\label{12114}
	\rP_{\bs}(\bs \notin \mathcal{S}_{\rm SD}|\bH)=\Gamma(2n_rT,\rho R_{\rm SD}^2)
	\end{align}
	where $\Gamma(n,x)$ is the normalized upper incomplete gamma function. Using asymptotic behaviour of $\Gamma(n,x)$ for high SNRs \cite{tmm}, (\ref{12114}) can be written as
	\begin{align}\label{12115}
	\rP_{\bs}(\bs \notin \mathcal{S}_{\rm SD}|\bH)=\frac{1}{(2n_rT-1)!}\left(\rho R_{\rm SD}^2\right)^{2n_rT-1}e^{-\rho R_{\rm SD}^2 }.
	\end{align}
	To achieve the DMT optimality, $\Delta \rP_{\rm SD}$ should satisfy $\Delta \rP_{\rm SD} \doteq \rho^{-d_{\rm ML}(r)}$. Therefore, according to (\ref{eq59}), (\ref{12111}) and (\ref{12115}) we have
	\begin{align}\label{12116}
	(\rho R_{\rm SD}^2)^{2n_rT-1}e^{-\rho R_{\rm SD}^2 }\doteq \rho^{-d_{\rm ML}(r)}.
	\end{align}
	Using the above equation, one can show that choosing  $R_{\rm SD}$ as 
	\beq
	R_{\rm SD}^2=d_{\rm ML}(r){\frac{\ln \rho}{\rho}},
	\eeq
	results in $\rP_{\bs}(\bs \notin \mathcal{S}_{\rm SD})\doteq \rho^{-d_{\rm ML}(r)}$ and, hence, the algorithm  achieves the DMT optimality. 
\end{proof}
 
According to Theorem 1, considering 
\beq
R_{\rm ZSD}^{(k)}=d_Q+\frac{\left(d_{\rm ML}(r) -d_{\rm MMSE}(r)\right)\ln\rho}{d_Q\rho_{{\rm MMSE}}^{(k)}}
\eeq 
and  
\beq
R_{\rm SD}^2=d_{\rm ML}(r){\frac{\ln \rho}{\rho}}
\eeq
as the radii, of the proposed algorithm guarantees
the DMT optimality.

In order to analyze the complexity behavior of the proposed algorithm, like most other works in the literature, we consider the number of visited nodes, denoted by $N$ \cite{sd9,sd10,jal11}. It is worth noting that, in the worst case, for the number of visited nodes we
have $N\doteq \rho^{rT}$. In other words, the number of visited nodes increases polynomialy with $\rho$. In \cite{jal11}, it
is shown that $\rho^{c(r)}$ represents the minimum required computational complexity to obtain the DMT optimal performance
for a conventional SD algorithm. It is also shown that $c(r)\neq0$ for $0<r<min(n_r,n_t)$, which implies that the computational
complexity for a DMT optimal method is an increasing function of $\rho$ at high SNR regime. The following theory shows that as the SNR tends to infinity, not only the
number of visited nodes in the proposed pruning algorithm tends to a constant number, but also visiting a single branch is sufficient to achieve the DMT optimality.

{\bf Theorem 2:} For the proposed method we have
\beq
\rP(N = 2n_tT)\dot{\geq} 1- \rho^{-d_{\rm MMSE}(r)}.
\eeq
\begin{proof}
		In order to analyze the behaviour of $N$ at high
	SNRs, we focus on calculating the probability of visiting $2n_tT$ nodes at high SNR, $\rP(N = 2n_tT)$
	\begin{align}\label{87871}
	&\rP(N = 2n_tT) = \rE_s\{\rP_\bs(N = 2n_tT)\},
	\end{align}
	where
	\begin{align}
	&\rP_\bs(N = 2n_tT) = \rP_{s_k}\left(\bigcap_{k=1}^{2n_tT}|\mathcal{S}^{(k)}|=1\right).
	\end{align}
	Since at each layer $\mathcal{S}^{(k)}=\mathcal{S}_{{\rm ZSD}}^{(k)}\bigcap \mathcal{S}_{{\rm SD}}^{(k)}$, we have $\rP_{s_k}(\bigcap_{k=1}^{2n_tT}|\mathcal{S}^{(k)}|=1)\geq \rP_{s_k}(\bigcap_{k=1}^{2n_tT}|\mathcal{S}_{{\rm ZSD}}^{(k)}|=1)$. Consequently,
	\begin{align}\label{87872}
	&\rP(N = 2n_tT) \geq \rP_{s_k}(\bigcap_{k=1}^{2n_tT}|\mathcal{S}_{{\rm ZSD}}^{(k)}|=1)=\nonumber\\
	& 1- \rP_{s_k}(\bigcup_{k=1}^{2n_tT}|\mathcal{S}_{{\rm ZSD}}^{(k)}|\neq 1)\geq 1-\sum_{k=1}^{2n_tT}\rP_{s_k}(|\mathcal{S}_{{\rm ZSD}}^{(k)}|\neq 1),
	\end{align}
	where the last inequality follows from the union bound. Based
	on the Bayes formula $\rP_{s_k}(|\mathcal{S}_{{\rm ZSD}}^{(k)}|
	\neq 1)$ can be expressed as
	\begin{align}\label{87873}
	&\rP_{s_k}(|\mathcal{S}_{{\rm ZSD}}^{(k)}|\neq 1)= \rP_{s_k}(|\mathcal{S}_{{\rm ZSD}}^{(k)}|\neq 1,R_{{\rm ZSD}}^{(k)}<2d_Q)\nonumber\\
	& +\rP_{s_k}(|\mathcal{S}_{{\rm ZSD}}^{(k)}|\neq 1,R_{{\rm ZSD}}^{(k)}>2d_Q)\nonumber\\&\leq \rP_{s_k}(|\mathcal{S}_{{\rm ZSD}}^{(k)}|\neq 1,R_{{\rm ZSD}}^{(k)}<2d_Q)+\rP(R_{{\rm ZSD}}^{(k)}>2d_Q),
	\end{align}
	It can be readily seen that if $R_{{\rm ZSD}}^{(k)} < 2d_Q$, merely the two events
	$|\mathcal{S}_{{\rm ZSD}}^{(k)}| = 1$ or $|\mathcal{S}_{{\rm ZSD}}^{(k)}| = 2$ can occur. Thus,
	\begin{align}\label{87874}
	&\rP_{s_k}(|\mathcal{S}_{{\rm ZSD}}^{(k)}|\neq 1,R_{{\rm ZSD}}^{(k)}<2d_Q)=\nonumber\\& \rP_{s_k}(|\mathcal{S}_{{\rm ZSD}}^{(k)}|= 2,R_{{\rm ZSD}}^{(k)}<2d_Q),
	\end{align}
	The joint events of $R_{{\rm ZSD}}^{(k)} < 2d_Q$ and $|\mathcal{S}_{{\rm ZSD}}^{(k)}| = 2$, is equivalent to the joint events of $R_{{\rm ZSD}}^{(k)} < 2d_Q$ and $\bigcup_{m=0}^{Q-2}\{s_{{\rm min,ZSD}}^{(k)}<c_m,s_{{\rm max,ZSD}}^{(k)}>c_{m+1}\}$. Hence, recalling $\bs_{{\rm min,ZSD}}$ and $\bs_{{\rm max,ZSD}}$, we have
	\begin{align}\label{87875}
	&\rP_{s_k}(|\mathcal{S}_{{\rm ZSD}}^{(k)}|= 2,R_{{\rm ZSD}}^{(k)}<2d_Q)=\nonumber\\&\sum_{m=0}^{Q-2}\rP_{s_k}(c_{m+1}-R_{{\rm ZSD}}^{(k)}<\tilde{y}_k<c_m+R_{{\rm ZSD}}^{(k)},R_{{\rm ZSD}}^{(k)}<2d_Q)
	\end{align}
	We assume that the $k$th transmitted symbol is $c_i$, i.e. $s_k = c_i$. Therefore,
	\begin{align}
	&\rP_{s_k}(|\mathcal{S}_{{\rm ZSD}}^{(k)}|= 2,R_{{\rm ZSD}}^{(k)}<2d_Q)=\sum_{i=0}^{Q-1}\sum_{m=0}^{Q-2}\nonumber\\&\rP(c_{m+1}-c_i-R_{{\rm ZSD}}^{(k)}<\tilde{w}_k<c_m-c_i+R_{{\rm ZSD}}^{(k)},R_{{\rm ZSD}}^{(k)}<2d_Q)
	\end{align}
	We have
	\begin{align}
	&\rP(c_{m+1}-c_i-R_{{\rm ZSD}}^{(k)}<\tilde{w}_k<c_m-c_i+R_{{\rm ZSD}}^{(k)},R_{{\rm ZSD}}^{(k)}<2d_Q) \nonumber\\&=\rP(|b_{m-i}|d_Q-\Delta R^{(k)}<\tilde{w}_k<|b_{m-i}|d_Q+\Delta R^{(k)},\Delta R^{(k)}<d_Q),
	\end{align}
	where $|b_{m-i}|=2(m-i)+1$. Since $\Delta R^{(k)}=\frac{\gamma}{d_Q\rho z_k}$, we obtain
	\begin{align}\label{87876}
	&\rP(|b_{m-i}|d_Q-\Delta R^{(k)}<\tilde{w}_k<|b_{m-i}|d_Q+\Delta R^{(k)},\Delta R^{(k)}<d_Q)\nonumber\\&=\int_{\frac{\gamma}{d^2_Q\rho}}^{\infty}\int_{|b_{m-i}|d_Q-\frac{\gamma}{d_Q\rho z_k}}^{|b_{m-i}|d_Q+\frac{\gamma}{d_Q\rho z_k}}f_{\tilde{W}_k}(\tilde{w}_k)f_{Z_k}(z_k)d\tilde{w}_k dz_k.
	\end{align}
	Using equation (\ref{87871})-(\ref{87876}), we have
	\begin{align}
	&\rP(N=2n_tT)\geq 1-\nonumber\\&\Big(\frac{1}{Q}\sum_{i=0}^{Q-1}\sum_{m=0}^{Q-2}\int_{\frac{\gamma }{d_Q^2\rho}}^{\infty}\int_{|b_{m-i}|d_Q-\frac{\gamma}{d_Q\rho z_k}}^{|b_{m-i}|d_Q+\frac{\gamma}{d_Q\rho z_k}}f_{\tilde{W}_k}(\tilde{w}_k)f_{Z_k}(z_k)d\tilde{w}_k dz_k\nonumber\\&+\rP(R_{{\rm ZSD}}^{(k)}>2d_Q)\Big).
	\end{align}
	According to the symmetry of Gaussian pdf, we have
	\begin{align}\label{antegral22}
	&\frac{1}{Q}\sum_{i=0}^{Q-1}\sum_{m=0}^{Q-2}\int_{\frac{\gamma }{d_Q^2\rho}}^{\infty}\int_{|b_{m-i}|d_Q-\frac{\gamma}{d_Q\rho z_k}}^{|b_{m-i}|d_Q+\frac{\gamma}{d_Q\rho z_k}}f_{\tilde{W}_k}(\tilde{w}_k)f_{Z_k}(z_k)d\tilde{w}_k dz_k=\nonumber\\&\frac{1}{Q}\sum_{i=0}^{Q-1}\sum_{m=0}^{Q-2}\int_{\frac{\gamma }{d_Q^2\rho}}^{\infty}\mathcal{Q}\left(|b_{m-i}|d_Q\sqrt{\rho z_k}-\frac{\gamma}{d_Q\sqrt{\rho z_k}}\right)f_{Z_k}(z_k)dz_k-\nonumber\\&\int_{\frac{\gamma }{d_Q^2\rho}}^{\infty}\mathcal{Q}\left(|b_{m-i}|d_Q\sqrt{\rho z_k}+\frac{\gamma}{d_Q\sqrt{\rho z_k}}\right)f_{Z_k}(z_k)dz_k).
	\end{align}
	Since, $d^2_{Q}\doteq \rho^{-\frac{r}{n_t}}$, considering the $\gamma$ of theorem 1, it is straight forward to show that, 
	\begin{align}
	&\int_{\frac{\gamma}{d_Q^2\rho}}^\infty \mathcal{Q}\left(|b_{m-i}|d_Q\sqrt{\rho z_k}+\frac{\gamma}{d_Q\sqrt{\rho z_k}}\right)f_{Z_k}(z_k)dz_k\nonumber \\& \leq \rho^{-d_{\rm ML}(r)}.
	\end{align}
	Now, we calculate the first integral of (\ref{antegral22}). Using Lemma 1 and similar to what we discussed in the previous section we have
	\begin{align}
	&\int_{\frac{\gamma}{d_Q^2\rho}}^\infty \mathcal{Q}\left(|b_{m-i}|d_Q\sqrt{\rho z_k}-\frac{\gamma}{d_Q\sqrt{\rho z_k}}\right)f_{Z_k}(z_k)dz_k \nonumber \\
	& \leq \int_{\frac{\gamma}{d_Q^2\rho}}^\infty \mathcal{Q}\left(|b_{m-i}|d_Q\sqrt{\rho t_k}-\frac{\gamma}{d_Q\sqrt{\rho t_k}}\right)f_{T_k}(t_k)dt_k,
	\end{align}
	where $T_k \sim \mathcal{X}_{2(n_r - n_t +1)}^2$. Therefore, according to $\mathcal{Q}(x)\leq \frac12 e^{-\frac{x^2}{2}}$ and Laplace theorem one writes
	\begin{align} \label{alakiIII}
	&\int_{\frac{\gamma}{d_Q^2\rho}}^\infty \mathcal{Q}\left(|b_{m-i}|d_Q\sqrt{\rho t_k}-\frac{\gamma}{d_Q\sqrt{\rho t_k}}\right)f_{T_k}(t_k)dt_k\leq \nonumber \\
	&\frac12 e^{{\gamma}|b_{m-i}|} \int_{\frac{\gamma}{d_Q^2\rho}}^{\infty}e^{-\frac{\rho}{2}(|b_{m-i}|^2d_Q^2t_k+\frac{\gamma^2}{d_Q^2\rho^2t_k})}f_{T_k}(t_k)dt_k\dot{\leq}\nonumber\\& \sqrt{\frac{2\pi\gamma}{\rho^2d_Q^4|b_{m-i}|^3}}\left(\frac{\gamma}{|b_{m-i}|d_Q^2\rho} \right)^{n_r-n_t}.
	\end{align}
	
	For the second term on the right hand side of (\ref{87873}), we have 
	\begin{align}
	&\rP(R_{{\rm ZSD}}^{(k)}>2d_Q)=\rP(\Delta R^{(k)}>d_Q)\nonumber\\&\rP(\rho_{{\rm MMSE}}^{(k)}<\frac{\gamma}{d_Q^2}).
	\end{align}
	According to the outage probability of the MMSE method, one can show that
	\beq\label{6565}
	\rP(\rho_{{\rm MMSE}}^{(k)}<\frac{\gamma}{d_Q^2})\doteq\rho^{-d_{\rm MMSE}(r)},
	\eeq
	where $d_{\rm MMSE}(r)=(n_r-n_t+1)(1-\frac{r}{n_t})$. Using equations (\ref{87872}), (\ref{alakiIII}) and (\ref{6565}), we  have
	\begin{align}\label{alakiV}
	&\rP\left(N=2n_tT\right) \dot{\geq} 1 - \nonumber\\& \Big(\frac{\rho^{-d_{\rm MMSE}(r)}}{Q}\sum_{i=0}^{Q-1}\sum_{m=0}^{Q-2}\frac{1}{|b_{m-i}|^{n_r-n_t+\frac32}}+\rho^{-d_{\rm MMSE}(r)}\Big).
	\end{align}
	Since the modulation order, $Q=\rho^r$ can be infinite, we have to show that the following series is bounded;
	\beq
	\mathcal{K} \triangleq \sum_{i=0}^{Q-1}\sum_{m=0}^{Q-2} \frac{1}{|b_{m-i}|^{n_r-n_t+\frac32}}.
	\eeq
	It should be noted that the maximum value of the above series is obtained when $n_r = n_t$. In this case
	\beq
	\mathcal{K}\leq \sum_{i=0}^{Q-1}\sum_{m=0}^{Q-2} \frac{1}{|2(m-i)+1|^\frac32}.
	\eeq
	By changing the variable $m-i=l$ we get
	\begin{align} 
	\mathcal{K} \leq \sum_{m=0}^{Q-2}\sum_{l=m}^{m-Q+1} \frac{1}{|2l+1|^\frac32} \leq \sum_{m=0}^{Q-1} \frac{1}{|2(m-Q+1)+1|^\frac32}.
	\end{align}
	For $Q\rightarrow\infty$, the above equation yields
	\begin{align}\label{alakiVI}
	\mathcal{K} \leq\sum_{m=0}^\infty \frac{1}{m^{\frac32}} = \xi(\frac32),
	\end{align}
	where $\xi(\cdot)$ is the Riemann zeta function and it is bounded \cite{zetaf}. Now using (\ref{alakiV}) and (\ref{alakiVI}) we get
	\begin{align}
	&\rP\left(N=2n_tT\right) \dot{\leq} Q-(Q-1)\times\nonumber \\&\left( 1-\frac{1}{Q}\xi(\frac32)\rho^{-d_{\rm MMSE}(r)}-\rho^{-d_{\rm MMSE}(r)}\right)\nonumber \\
	&\dot{\leq} 1 - \rho^{-d_{\rm MMSE}(r)}.
	\end{align}
\end{proof}
 
Theorem 2 implies that, with a probability that asymptotically tends to  one, the proposed algorithm visits a single branch at high SNR regime. 
\section{simulation Results} \label{simres}
In this section, some numerical examples are presented as metrics for measuring the performance and the complexity of the proposed method in various SNRs. The MIMO channels are considered to be i.i.d complex Gaussian with zero mean and unit variance. Two simulation types are presented. The first type considers the variable rate scenario ($r\neq0$) and in the second type, the fixed rate scenario ($r=0$) is assessed.
\subsection{Variable Rate ($r\neq0$)} 
To substantiate the results of Theorems 1 and 2, Fig. \ref{performance4_5} and Fig. \ref{cardcoded} are given where $\Delta \rP$ and $\rP(N = 2n_tT)$ are shown versus SNR, respectively. These figures are obtained for a system with two transmit and receive antennas adopting Golden code \cite{sd3} with different multiplexing gains of $r=0.8$, $r=1$, $1.2$ and $r=1.4$. According to Theorem 1, we have simulated the following upper bound for the performance gap 
\beq
\Delta \rP = \rP\left( \bs \notin \mathcal{S}_{\rm ZSD}\right) + \rP \left( \bs \notin \mathcal{S}_{\rm SD}   \right).
\eeq   
As it can be observed, from Fig. \ref{performance4_5}, the proposed method is DMT optimal  
which corroborates with the results of Theorem 1. 
To investigate the behaviour of the computational complexity of the proposed method at high SNR regime, $\rP(N = 2n_tT)$ is calculated in Fig. \ref{cardcoded}. It can be seen that at high SNR a single branch, or equivalently a single node per layer is visited which agrees with Theorem 2. 
\subsection{Fixed Rate ($r=0$)}
Figs. \ref{asarantenper} and \ref{asarantencomp}, compare the performance and the complexity of the proposed pruning method with that of some other pruning methods which achieve a relatively low complexity and a near ML performance. This simulated example is for a fixed rate, $r=0$, and a Golden coded MIMO system with two transmit and receive antennas with 64-QAM. In example, the following methods are simulated for comparison
\begin{itemize}
	\item Increased radius search sphere decoding (IRS-SD)\cite{sd9}.
	\item Performance achieving reduced complexity sphere decoding (PARC-SD) \cite{sd18}.
	\item Probabilistic tree pruning SD with inter-search
	radius control (PTP-SD+ISRC) \cite{2010}.
	\item Threshold pruning SD\cite{2013}.
\end{itemize}
It can be seen that, although all the methods achieve near ML performance, the proposed pruning method benefits from a lower computational complexity in terms of average number of flops.
\begin{figure}
	\centering
	
	\begin{tikzpicture}[scale=.9,every label/.style]
	\begin{semilogyaxis}[axis background/.style={
		shade,top color=white!10,bottom color=white},
	legend style={fill=white},
	legend style={font=\tiny},
	legend style={at={(0.03,0.03)},
		anchor=south west},
	xmin = 15, xmax = 68,
	ymin = 0.001,
	xlabel={SNR (dB)},
	ylabel={$\Delta\rP$},
	grid=major,
	legend entries={$r=0.8$,$r=1$,$r=1.2$,$r=1.4$},
	]

	\addplot+[thick, mark=otimes*,color=red]  coordinates{                                     
		(12.04+3,0.3541) (27.09+3,0.01056) (42.14+3,0.000115) (57.2+3,0.000001)
	};
	\addplot+[thick, mark=diamond*,color = green]  coordinates{                                     
		(9.031+3,0.8836) (21.07+3,0.2867) (33.11+3,0.05408) (45.15+3,0.00782) (57.2+3,0.00097)
	};
	\addplot+[thick,mark=triangle*,color=blue]  coordinates{                                 
		(7.024+3,0.991) (17.06+3,0.6515) (27.09+3,0.268) (37.13+3,0.0968) (47.16+3,0.0265) (57.2+3,0.0079) (67.23+3,0.0019)
	};
	\addplot+[thick,mark=square*,color=black]  coordinates{                                 
		(5.591+3,0.999) (14.19+3,0.9453) (22.79+3,0.635) (31.39+3,0.387) (39.99+3,0.207) (48.59+3,0.0975) (57.2+3,0.0461) (65.8+3,0.0182)
	};
	
	\node [label={[label distance=2.9cm]0.9:$\simeq\rho^{-d_{\rm ML}(1.4)}$}]{};
	\node [label={[label distance=2.1cm]-1.9:$\xrightarrow{\hspace{0.5cm}}$}]{};
	
	\node [label={[label distance=5.1cm]-4.2:$\simeq\rho^{-d_{\rm ML}(1.2)}$}]{};
	\node [label={[label distance=2.9cm]-10.7:$\xrightarrow{\hspace{1.9cm}}$}]{};
	
	\node [label={[label distance=2.9cm]-44.9:$\simeq\rho^{-d_{\rm ML}(0.8)}$}]{};
	\node [label={[label distance=5.6cm]-9.2:$\simeq\rho^{-d_{\rm ML}(1)}$}]{};
	\node [label={[label distance=2.9cm]-20.7:$\xrightarrow{\hspace{2.5cm}}$}]{};
	
	\addplot+[thick, mark=-,color=red, dashed]  coordinates{                                     
		(12.04+3,0.9541) (27.09+3,0.01056) (42.14+3,0.000115) (57.2+3,0.000001)
	};
	\addplot+[thick, mark=-,color = green, dashed]  coordinates{                                     
		(9.031+3,4.8836) (21.07+3,0.5867) (33.11+3,0.07408) (45.15+3,0.00782) (57.2+3,0.00097)
	};
	\addplot+[thick,mark=-,color=blue, dashed]  coordinates{                                 
		(7.024+3,6.991) (17.06+3,1.7515) (27.09+3,0.468) (37.13+3,0.128) (47.16+3,0.03365) (57.2+3,0.0079) (67.23+3,0.0019)
	};
	\addplot+[thick,mark=-,color=black, dashed]  coordinates{                                 
		(5.591+3,3.999) (14.19+3,1.9653) (22.79+3,1.005) (31.39+3,0.507) (39.99+3,0.247) (48.59+3,0.105) (57.2+3,0.0461) (65.8+3,0.0182)
	};;

	\end{semilogyaxis}grid=major,

	\end{tikzpicture}

	\caption{The perfomance gap between the proposed method and the ML decoding for a $2\times 2$ Golden coded MIMO system  for diverse multiplexing gain}
	\label{performance4_5}
\end{figure}
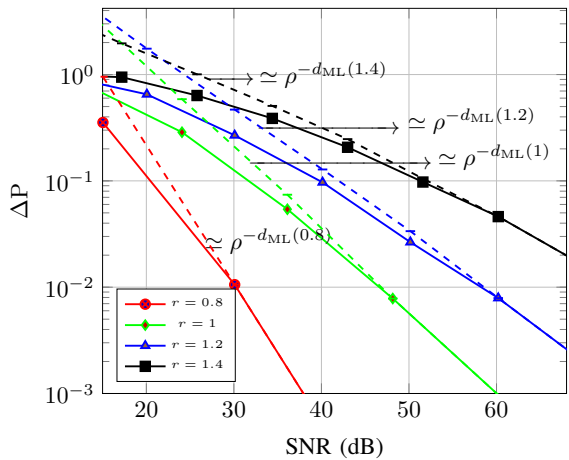 

\begin{figure}
	\centering
\begin{tikzpicture}[scale=.9]
\begin{axis}[axis background/.style={
	shade,top color=white!10,bottom color=white},
legend style={fill=white},
legend style={font=\tiny},
legend style={at={(.95,0.05)},
	anchor=south east},
xmin = 20, xmax = 102,
ymin = 0.566,ymax=1,
xlabel={SNR(dB)},
ylabel={$\rP(N = 2n_tT)$},
grid=major,
legend entries={$r=0.8$,$r=1$,$r=1.2$,$r=1.4$},
]

\addplot+[thick, mark=otimes*,color=red] 
table[row sep=crcr]{              
	15.0515  0.665 \\
	30.1030  0.892  \\
	45.1545  0.979  \\
	60.2060 0.994 \\
	75.2575  0.999   \\
	90.3090  1  \\
	105.3605 1\\
	120.4120 1\\
	135.4635 1\\
};

\addplot+[thick, mark=diamond*,color = green]  
table[row sep=crcr]{              
	12.0412  0.615 \\
	24.0824  0.74  \\
	36.1236   0.914  \\
	48.1648 0.97 \\
	60.2060  0.995   \\
	72.2472  0.997  \\
	84.2884 1\\
	96.3296 1\\
	108.3708 1\\
};

\addplot+[thick,mark=triangle*,color=blue]  
table[row sep=crcr]{              
	20.0687  0.566 \\
	30.1030  0.739  \\
	40.1373 0.868 \\
	50.1717  0.942   \\
	60.2060  0.969 \\
	70.2403 0.992\\
	80.2747 0.998\\
	90.3090 1\\
};

\addplot+[thick,mark=square*,color=black]  
table[row sep=crcr]{              
	17.2017  0.416  \\
	25.8026  0.502  \\
	34.4034 0.625\\
	43.0043  0.737  \\
	51.6051  0.847  \\
	60.2060 0.913\\
	68.8069 0.941\\
	77.4077 0.965\\
	86.0086 0.981\\
	94.6094 0.991\\
	103.2103 1\\
};

\end{axis}grid=major,

\end{tikzpicture}

	\caption{The cardinality of visited nodes for a $2\times 2$ Golden coded MIMO system  for diverse multiplexing gain}
	\label{cardcoded}
\end{figure}
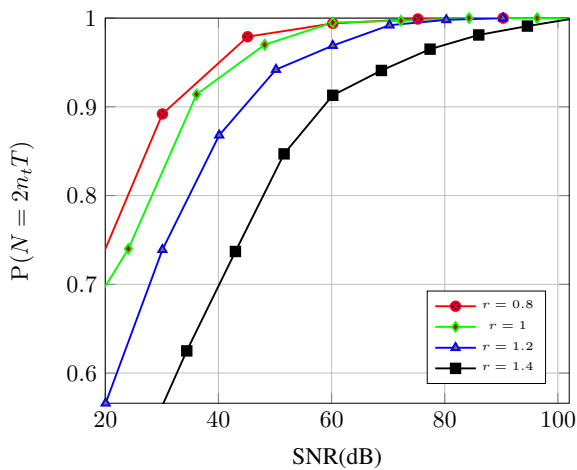 
\begin{figure}
	\centering
	\begin{tikzpicture} [scale = .9]
	
	\begin{semilogyaxis}[
	axis background/.style={
		shade,top color=white!10,bottom color=white},
	legend style={draw=darkgray!60!black,fill=white,legend cell align=left},
	legend style={fill=white},
	legend style={font=\tiny},
	legend style={at={(0.03,0.03)},
		xmin = 0, xmax = 30,
		ymin = 0.0002, 
		anchor=south west},
	xlabel={SNR (dB)},
	ylabel={Symbol error probability},
	grid=major,
	legend entries={ML,PARC-SD\cite{sd18},Proposed Method,PTP-SD+ISRC\cite{2010},Threshold Pruning SD\cite{2013}},
	]

	\addplot [
	thick,mark=square,color= black
	]
	table[row sep=crcr]{
		0  0.8245 \\
		5 0.7060 \\
		10 0.4640  \\
		15  0.1600\\
		20 0.03174\\
		25  0.00492\\
		30 0.00023\\
	};
	\addplot [
	thick,mark=star,color= green,dashed
	]
	table[row sep=crcr]{
		0  0.8560   \\
		5   0.7060 \\
		10 0.4640  \\
		15  0.1600\\
		20 0.03174\\
		25  0.00492\\
		30 0.00023\\
	};
	
	\addplot [
	thick,
	color=red,
	solid,
	mark=o,
	mark options={solid}
	]
	table[row sep=crcr]{
		0  0.8455  \\
		5   0.7095\\
		10 0.4640  \\
		15  0.1600\\
		20 0.03174\\
		25  0.00492\\
		30 0.00023\\
	};
	\addplot [
	thick,
	color=blue,
	dotted,
	mark=asterisk,
	mark options={dashed}
	]
	table[row sep=crcr]{
		0 0.8245  \\
		5   0.7160 \\
		10 0.4740  \\
		15  0.1640\\
		20 0.03274\\
		25  0.00532\\
		30 0.00025\\
	};
	\addplot [
	thick,
	color=cyan,
	solid,
	mark=diamond,
	]
	table[row sep=crcr]{
		0 0.8470  \\
		5 0.7075  \\
		10 0.4640  \\
		15  0.1600\\
		20 0.03174\\
		25  0.00492\\
		30 0.00023\\
	};
	
	\end{semilogyaxis}
	\end{tikzpicture}%
	\caption{The symbol error rate comparison of different SD detectors for the 64-QAM, $2\times 2$ Golden Coded ($T=2$) MIMO system.}
	\label{asarantenper}
\end{figure}
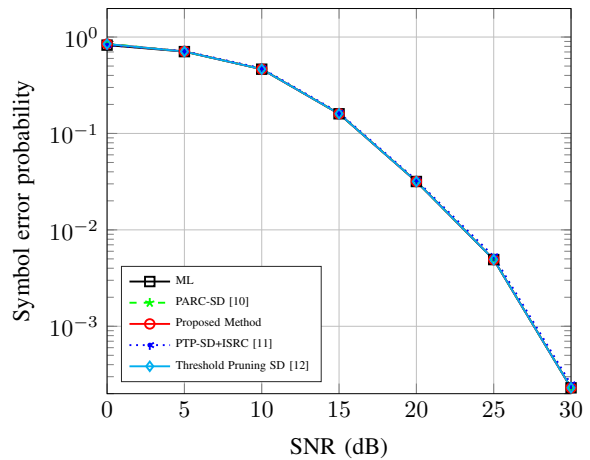
\begin{figure}
	\centering
	\begin{tikzpicture}[scale =.9]
	
	\begin{axis}[%
	axis background/.style={
		shade,top color=white!10,bottom color=white},
	legend style={fill=white},
	legend style={font=\tiny},
	legend style={at={(.55,0.67)},
		anchor=south west},
	xmin = 0, xmax = 30,
	ymin = 0, ymax = 1000000,
	xlabel={SNR (dB)},
	ylabel={Average number of flops},
	grid=major,
	legend entries={IRS-SD\cite{sd9},PARC-SD\cite{sd18},Proposed Method,PTP-SD+ISRC\cite{2010},Threshold Pruning SD\cite{2013}},
	]

	\addplot [
	thick,mark=square,color= black
	]
	table[row sep=crcr]{              
		0  1672900  \\
		5  853300  \\
		10   494500   \\
		15 251800  \\
		20  79500   \\
		25   12500  \\
		30 2200\\
	};
	\addplot [
	thick,mark=star,color= green,dashed
	]                  
	table[row sep=crcr]{	
		0  0   \\
		5   591700     \\
		10  355000      \\
		15   130700    \\
		20   35500    \\
		25  13000 \\
		30    3400\\
	};
	
	\addplot [
	thick,
	color=red,
	solid,
	mark=o,
	mark options={solid}
	]                              
	table[row sep=crcr]{   	
		0 	700 \\
		5  	310500 \\
		10  181830\\
		15  71640\\
		20  31150   \\
		25  12600 \\
		30  2700\\
	};
	\addplot [
	thick,
	color=blue,
	dotted,
	mark=asterisk,
	mark options={dashed}
	]                           
	table[row sep=crcr]{
		0 942300    \\
		5   487700 \\
		10   252100  \\
		15   85800   \\
		20 	 36100\\
		25   18000\\
		30   5300\\
	};
	\addplot [
	thick,
	color=cyan,
	solid,
	mark=diamond,
	]                               
	table[row sep=crcr]{	
		0 516100 \\
		5   533700  \\
		10  414900    \\
		15    217100 \\
		20    69700\\
		25  15800  \\
		30   3900\\
	};
	\end{axis}
	\end{tikzpicture}%
	\caption{The average number of flops comparison of different SD detectors for the 64-QAM, $2\times 2$ Golden Coded ($T=2$) MIMO system.}
	\label{asarantencomp}
\end{figure}
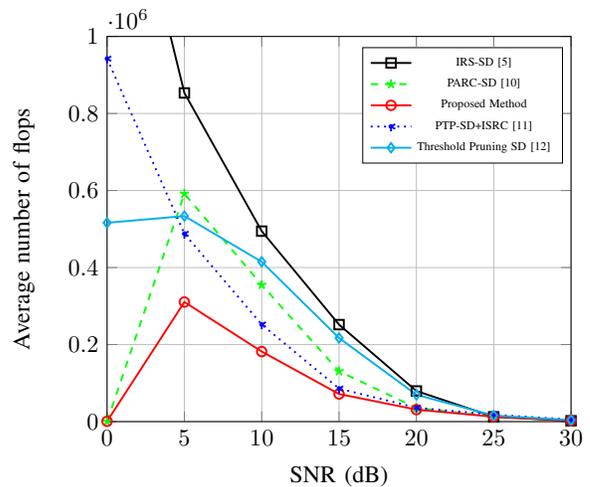
\section{Conclusion} \label{conclusion}
In this paper a tree pruning method  that intersects some zero-spheres with the hypersphere of the SD algorithm is proposed. The radii corresponding to the zero-spheres and the SD algorithm are designed to guarantee the DMT optimality. Beyond the DMT optimality, we show that the proposed algorithm only needs to visit a single node at high SNR regime and high transmission rates. This result shows that, unlike the conventional SD algorithms which are shown to have a polynomial complexity behavior at high SNR regime and high transmission rates, the number of visited nodes of the proposed method becomes constant (exactly one node) as the SNR increases.
\section*{Appendix A}
\section*{Proof of Lemma 1}
In \cite{zeroforcing} it is proven that $\rho_{{\rm MMSE}}^{(k)}\geq \rho_{{\rm ZF}}^{(k)}$ where $\rho_{{\rm ZF}}$ is the zero forcing (ZF) SNR. Now, we prove that there exists an $\alpha$ such that $ \rho_{{\rm MMSE}}^{(k)}\leq\rho_{{\rm ZF}}^{(k)}+\alpha$. It is known that for the MMSE SNR 
\beq
\rho_{{\rm MMSE}}^{(k)} = \frac{\rho}{[(\bH^\rT\bH+\rho\bI)^{-1}]_{kk}}-1.
\eeq
Since, $\bH= \mathcal{H}\bR$ we have $\bH^\rT\bH= \bR^\rT\bB\bR$ where $\bB = \mathcal{H}^\rT\mathcal{H}$ is a positive definite matrix. By defining $\bA = (\bR^\rT)^{-1}\bR^{-1}$ we get
\beq
\rho_{{\rm MMSE}}^{(k)} = \frac{\rho}{\be_k^\rT\bR^{-1}(\bB+\rho^{-1}\bA)^{-1}(\bR^\rT)^{-1}\be_k }-1,
\eeq
for a positive definite matrix $\bA \leq \lambda_{\max} \bI$ where $\lambda_{\max}$ is the maximum eigen value of the matrix $\bA$, which yields
\beq
\rho_{{\rm MMSE}}^{(k)} \leq \frac{\rho}{\be_k^\rT\bR^{-1}(\bB+\rho^{-1}\lambda_{\max}\bI)^{-1}(\bR^\rT)^{-1}\be_k }-1.
\eeq
Now according to the eigen value decomposition of the positive definite matrix $\bB$ we have $\bB = \bQ^\rT\boldsymbol{\Lambda}_{B}\bQ$ where $\bQ$ is the eigen vector matrix and $\boldsymbol{\Lambda}_\bB$ is the eigen value matrix of $\bB$. Therefore
\beq
(\bB+\rho^{-1}\lambda_{\max}\bI)^{-1} = \bQ^{-1}(\boldsymbol{\Lambda}_\bB+ \Delta\bI)^{-1} (\bQ^\rT)^{-1},
\eeq
where $\Delta\triangleq\rho^{-1}\lambda_{\max}$. Hence, by defining $\bl_k = (\bQ^\rT)^{-1}(\bR^\rT)^{-1}\be_k$, we get
\begin{align}
\rho_{{\rm MMSE}}^{(k)} &\leq\frac{\rho}{\be_k^\rT\bR^{-1}\bQ^{-1}(\boldsymbol{\Lambda}_\bB+ \Delta\bI)^{-1} (\bQ^\rT)^{-1}(\bR^\rT)^{-1}\be_k}-1\nonumber\\ 
&= \frac{\rho}{\sum_{i=1}^{2n_tT}\bl_k^2(i)\frac{1}{\lambda_{\bB}(i)+\Delta}}-1,
\end{align}
where $\lambda_{\bB}(i)$s are the eigen values of the matrix $\bB$. Therefore, one can show that
\beq
\rho_{{\rm ZF}}^{(k)} = \frac{\rho}{\sum_{i=1}^{2n_tT}\frac{\bl_k^2(i)}{\lambda_{\bB}(i)}}.
\eeq
As we mentioned previously, our aim is to find $\alpha$ such that $\rho_{{\rm ZF}}^{(k)}-\rho_{{\rm MMSE}}^{(k)} +\alpha \geq0$. It is straightforward to show that in order for $\rho_{{\rm ZF}}^{(k)}-\rho_{{\rm MMSE}}^{(k)} +\alpha \geq0$ to be  satisfied, it is sufficient that the following equation be positive
\beq
\sum_i \frac{-\rho \bl_k^2(i)}{\lambda_{\bB}(i)} +\frac{\rho \bl_k^2(i)}{\lambda_{\bB}(i)+\Delta} + \frac{(\alpha+1)\bl_k^4(i)}{\lambda_{\bB}(i)(\lambda_{\bB}(i)+\Delta)}\geq 0,
\eeq
which leads to the following equality
\beq
\alpha \geq \frac{\lambda_{\max}\sum_{i}\bl_k^2(i)}{\sum_i \bl_k^4(i)}-1.
\eeq 
According to the Schwarz  inequality  we have 
\beq
\frac{1}{2n_tT}\left(\sum_{i=1}^{2n_tT}\bl_k^2(i)\right)^2 \leq \sum_{i=1}^{2n_tT}\bl_k^4(i). 
\eeq 
Hence, it is sufficient to choose 
\beq
\alpha \geq \frac{2n_tT\lambda_{\max}}{\|\bR^{-1}\be_k \|^2}-1.
\eeq
According to
\beq
\frac{\lambda_{\max}}{\|\bR^{-1}\be_k \|^2} >1,
\eeq
and the fact that $\rho_{{\rm MMSE}}^{(k)} \geq \rho_{{\rm ZF}}^{(k)}$, we get
\beq \label{jaleb}
\rho_{{\rm ZF}}^{(k)} \leq \rho_{{\rm MMSE}}^{(k)} \leq \rho_{{\rm ZF}}^{(k)} + 2n_tT-1.
\eeq
Therefore, we have
\beq
\int_0^\infty g(z_k)f_{Z_k}(z_k)dz_k = \int_0^\infty g(z_k)\rP(z_k\leq Z_k\leq z_k+dz_k)dz_k.
\eeq
According to (\ref{jaleb}), we have $\rho x_k \leq \rho z_k \leq \rho x_k + 2n_tT-1$. Thus,
\begin{align}
&\rP(z_k\leq Z_k\leq z_k+dz_k) \nonumber \\ 
&\leq \rP(z_k-\frac{2n_tT-1}{\rho}\leq X_k\leq z_k+dz_k) \nonumber \\
& \leq \rP(z_k\leq X_k\leq z_k+dz_k).
\end{align}


\begin{thebibliography}{1}
	\bibitem{new1}
	 K. Yang and S. Tsai, ``Maximum Likelihood and Soft Input Soft Output MIMO Detection at a Reduced Complexity," in IEEE Transactions on Vehicular Technology, vol. 67, no. 12, pp. 12389-12393, 2018.
	\bibitem{new2}
	M. Mohammadkarimi, et.al. ``Deep Learning-Based Sphere Decoding," in IEEE Transactions on Wireless Communications, vol. 18, no. 9, pp. 4368-4378, 2019.
	\bibitem{50yearsmimo} 
	S. Yang and L. Hanzo,  ``Fifty years of MIMO detection: The road to large-scale MIMOs," {\it IEEE Communications Surveys and Tutorials}, vol. 17, no. 4, pp. 1941-88, Sep. 2015.
	\bibitem{sd10} 
	J.~Jald\'{e}n and B.~ Otterste. ``On the complexity of sphere
	decoding in digital communications."  {\it IEEE Trans. on Signal
		Processing}, vol.~53, no.~4
	pp.~1474-1484, Apr. 2005.
	
	\bibitem{sd9} 
	B. Hassibi and H. Vikalo, ``On the sphere-decoding algorithm I. expected complexity," IEEE Transactions on Signal Processing, vol. 53, pp. 2806–2818, 2005.
	
	\bibitem{Pauli}
	V. Pauli and L. Lampe, ``On the complexity of sphere decoding for
	differential detection," {\it IEEE Trans. Inf. Theory}, vol. 53, no. 4, pp.
	1595–1603, Apr. 2007.
	
	\bibitem{tsebook} 
	D. Tse and P. Viswanath, {\it Fundamentals of wireless communications}. Cambridge, 2005.
	
	
	\bibitem{jal11} 
	J. Jalden,  P. Elia ``Sphere decoding complexity exponent for decoding full-rate codes over the quasi-static MIMO channel" {\it IEEE Transactions on Information Theory}, vol. 58, no. 9, pp. 5785-5803, Sep. 2012.
	\bibitem{sd3} 
	E. Viterbo and J. Boutros, ``A universal lattice code decoder for fading channels," IEEE Trans. Inf. Theory, vol. 45, no. 5, pp. 1639–1642, Jul. 1999.
	\bibitem{sd18} 
	M. Neinavaie, and M. Derakhtian, `` ML performance achieving algorithm with the zero-forcing complexity at high SNR regime,'' {\it IEEE Transactions on Wireless Communications,} vol.~15, no.~7, pp.~4651-4659, jul. 2016.
	\bibitem{2010} 
	B. Shim and I. Kang, ``On further reduction of complexity in tree pruning based sphere search" {\it IEEE transactions on Communications}, vol. 58, no. 2, Feb. 2010.
	\bibitem{2013} 
	T. Cui,  S. Han,  and  C. Tellambura, ``Probability-distribution-based node pruning for sphere decoding" {\it IEEE Transactions on Vehicular Technology}, vol. 62, no. 4, pp. 1586-1596, May 2013.
\bibitem{zeroforcing}
Yi Jiang, Varanasi M. K.  and Jian Li, ``Performance analysis of ZF and MMSE equalizers for MIMO systems: an in-depth study of the high SNR regime , " {\it IEEE Transaction on Information Theory }, vol. 57, no. 4, pp. 2008 - 2026 , April 2011.
\bibitem{jerry}
 N. C. Giri, {\it Multivariate statistical inference.} Academic Press, 2014.
\bibitem{1994lap}
A. Azevedo-Filho, and R. D. Shachter. ``Laplace's method approximations for probabilistic inference in belief networks with continuous variables." In Uncertainty Proceedings 1994, pp. 28-36. Morgan Kaufmann, 1994.
\bibitem{tmm}
N. M.Temme, ``Uniform asymptotic expansions of the incomplete gamma functions and the incomplete beta function," {\it Mathematics of Computation}, vol. 29, no. 132, pp.~1109-1114, 1975.
\bibitem{zetaf}
H. M. Srivastava  and J. Choi, {\it Series associated with the zeta and related functions}. Vol. 530. Springer Science and Business Media, 2001.
\end{thebibliography}
\end{document}